\documentclass{ledger}

\hypersetup{pdfauthor={Hafner, Matthias; 
Dietl, Helmut}, pdftitle={Paper Title}}

\usepackage{booktabs} %%

\title{Impermanent Loss Conditions: An Analysis\\
of Decentralized Exchange Platforms}
\author{Matthias Hafner\thanks{University of Zurich, Center for Cryptoeconomics, and Swiss Economics} , Helmut Dietl\thanks{University of Zurich}}

\pretitle{
 \vspace{1em}
 \centering \selectfont CfC St. Moritz Academic Research Track 2024 \par
 \fontsize{24pt}{28pt}\selectfont} % Title is centered and at 24pt

\begin{document}

\maketitle

\thispagestyle{pagefirst}

\begin{abstract}
Decentralized exchanges are widely used platforms for trading crypto assets. The most common types work with automated market makers (AMM), allowing traders to exchange assets without needing to find matching counterparties. Thereby, traders exchange against asset reserves managed by smart contracts. These assets are provided by liquidity providers in exchange for a fee. Static analysis shows that small price changes in one of the assets can result in losses for liquidity providers. Despite the success of AMMs, it is claimed that liquidity providers often suffer losses. However, the literature does not adequately consider the dynamic effects of fees over time. Therefore, we investigate the impermanent loss problem in a dynamic setting using Monte Carlo simulations. Our findings indicate that price changes do not necessarily lead to losses. Fees paid by traders and arbitrageurs are equally important. In this respect, we can show that an arbitrage-friendly environment benefits the liquidity provider. Thus, we suggest that AMM developers should promote an arbitrage-friendly environment rather than trying to prevent arbitrage.

%AUTHOR: keywords are OK to show for Review article, will be hidden and added to metadata for publication
\end{abstract}

\section{Introduction}
Decentralized exchanges (DEXs) are essential for many blockchain applications as they enable users to access different services and investors to access various (crypto) assets. In comparison to centralized exchanges, DEXs are simple to access, do not require any customer identity verification, and can be seamlessly integrated into other decentralized financial (DeFi) services at almost no cost. The Block Research reports that the combined trading volume of all DEXs exceeded USD 1,000 billion in 2021.\cite{BlockResearch}

DEXs function without an orderbook. Traders are not matched, but rather exchange their assets against a liquidity pool. This liquidity pool is filled with assets by so-called liquidity providers, who receive fees in return for providing liquidity. The prices at which traders can exchange their assets against the liquidity pool are adjusted by automated market makers (AMMs). AMMs are simple mathematical formulas or more complex algorithms that determine quantities and prices for the assets in the liquidity pools. These formulas or algorithms guarantee that liquidity pools never run out of any of their assets, adjusting prices accordingly. If, for example, a liquidity pool consists of USDC and ETH and the number of ETH decreases, the AMM increases the price of ETH in USDC. AMMs are usually programmed as smart contracts on blockchains.

Due to these automated quantity and price adjustments, the total value of the assets in the liquidity pool can be lower than their value would have been had the liquidity providers kept these assets in their wallets. This difference is called impermanent loss and is described by liquidity providers’ profit function.\cite{AignerDhaliwal, Bancor} Many practitioners argue that impermanent loss is a major issue. As a result of this belief, many DEXs updated their protocols accordingly (e.g., Bancor’s Impermanent Loss Protection).\cite{Bancor} However, the growth and market share of Uniswap, the world’s largest DEX (which did not address the problem directly), does not indicate impermanent loss to be the most relevant issue for liquidity providers.\cite{BlockResearch}

The peer-reviewed literature on impermanent losses is limited. The earliest contributions are from Angeris et al., published in 2021.\cite{Angeris, Chitra} Assuming no fees, they formally analyze Uniswap V2 and show that the liquidity providers’ profits are negatively related to changes in market prices and that AMMs closely track the reference market price. Evans (2021) shows that these results also hold for generalized models.\cite{Evan} Our analysis builds on this analysis. In addition, there are some working papers focusing on impermanent loss. Angeris et. al (2022) compute the profit functions of liquidity providers in liquidity pools with constant function (automated) market makers.\cite{Chitra} Other authors demonstrate approximate hedging techniques.\cite{Clark, Fukashe, Milionis2} Aigner and Dhaliwal (2021) describe the risk profile of a liquidity provider and compute the impermanent loss function for Uniswap V2 and show that in a static setting, small changes in the relative market price between two assets result in an impermanent loss. In particular, they show that the impermanent loss is an inversely U-shaped function of the relative price changes of the underlying assets/tokens. In a similar setting, Labadie (2022) shows that the impermanent loss increases faster than linear and disappears after a price reversion.\cite{Labadie} Lehar and Parlour (2023) describe the impermanent loss as a function of fee gains and adverse rebalancing of the AMM. They show that liquidity provision is endogenously determined by this trade-off.\cite{Lehar} 

We build on the literature partly to get a better understanding of the profit dynamics of liquidity providers. In contrast to previous literature, however, we explicitly include accumulated trader and arbitrageur fees over time in the profit function of liquidity providers. We use an agent-based model that allows us to consider dynamic effects between participants over time without making too many restrictions regarding their interactions. We simulate the behavior of liquidity providers, traders, and arbitrageurs under various scenarios to compute the resulting profits and losses of liquidity providers based on a constant product AMM (see, e.g., Uniswap’s V2). This approach allows us to identify under which conditions liquidity providers will incur impermanent losses and how the inclusion of fees changes standard results. 

The remainder of the paper is structured as follows. Section \ref{sec:AMM_economics} introduces the basic economics of AMMs and the relevant literature. Section \ref{sec:AMM_ABM} describes the setup of the agent-based model and explains the simulation. Section \ref{sec:AMM_Results} presents and discusses the results. Section \ref{sec:AMM_Conclusion} concludes.

\newpage

\section{The Economics of Decentralized Exchanges} \label{sec:AMM_economics}
\begin{figure} [h]
\centering
\captionsetup{justification=centering}
\includegraphics[width=0.8\textwidth]{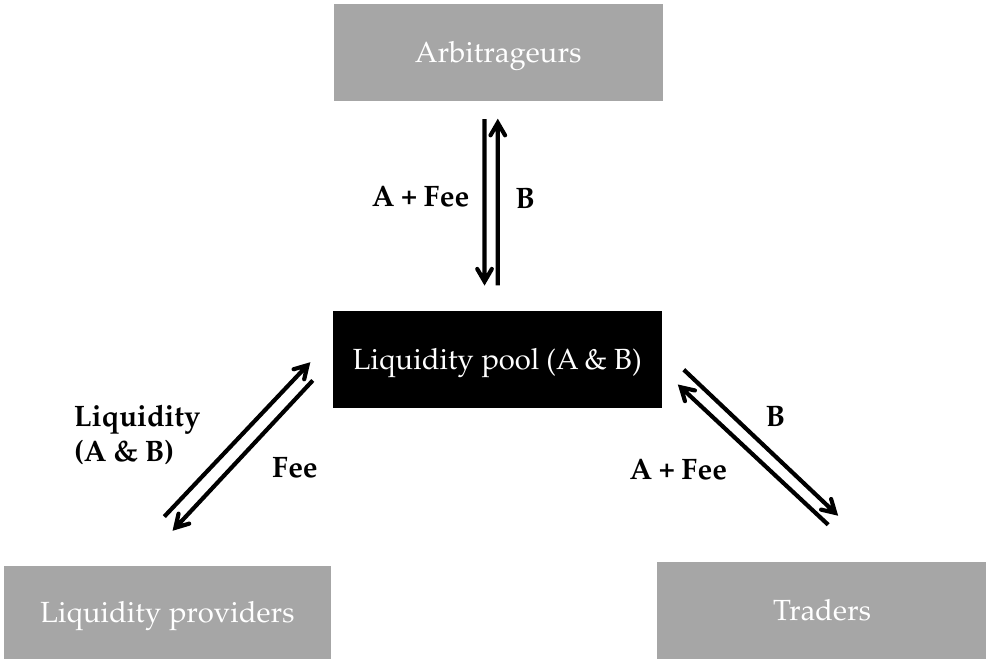}
 \caption{Illustration of the interactions of different market players in a DEX.}
 \label{graph4}
\end{figure}

As shown in Figure \ref{graph4}, DEXs are three-sided platforms. Liquidity providers interact with the platform by supplying assets to the liquidity pool in exchange for a fee. The liquidity pool depicted in Figure \ref{graph4} consists of two assets denoted A and B, for example, USDC and WETH. Traders can exchange one of the two assets for the other by paying a fee. Arbitrageurs also interact with the platform whenever they perceive price differences between the exchange rate of A and B at the DEX and that of other markets. If, for example, asset B is more valuable (relative to Asset A) at another exchange, arbitrageurs will buy B from the DEX by sending asset A and a fee to this DEX in exchange for asset B. At the same time, these arbitrageurs will sell B for A at another exchange to earn a risk-free profit.

\vspace{12pt}

Based on the amounts of the two assets/tokens in the liquidity pool (denoted by $R_A$ and $R_B$, respectively), the AMM determines the price at which traders and arbitrageurs can exchange assets with the liquidity pool. If traders or arbitrageurs send an amount of asset A (denoted by $\Delta_A$) to the liquidity pool they will receive an amount of asset/token B (denoted by $\Delta_B$) in return according to the fee and the price determined by the AMM. Of course, traders and arbitrageurs can also buy B by sending the required amount of A to the liquidity pool. The liquidity pool was filled with asset A and B by liquidity providers in advance. For this liquidity provision they receive fees from the DEX. 
Of particular interest is the AMM’s formula or algorithm that determines the exchange rates offered to traders and arbitrageurs. These algorithms adjust exchange rates based on the relative demand and supply of all assets in the liquidity pool. If there is more demand for asset B in relation to A, the AMM will increase the exchange rate. Traders and arbitrageurs will have to send more units of A to receive one unit of B. In addition, most AMMs set prices such that pools never run out of any of their assets/tokens. 
In this paper, we focus on AMMs that use a “constant product” function such as the one implemented in the Uniswap V2 protocol for the determination of the exchange rate. We focus on this type because it was not only the first, but currently, is also the most common type. Note, however, that there also exist other exchange rate formulas and algorithms, in particular for stable swaps, i.e., for asset pairs whose prices are (almost) identical.\cite{Egorov, Angeris, XU} Uniswap’s updated version (V3) is also based on the constant product function. It allows liquidity providers to define a price range at which the liquidity pool can offer trades. For simplicity, we did not model this additional feature.

In line with previous research we describe the two token/asset constant product function by 
\begin{equation}
R_A*R_B=k \label{eq:kfunc_main}
\end{equation}
where $k$ is a constant value and $R_i$ are the reserves, i.e., the actual amount of asset $i$ in the liquidity pool.\cite{Angeris, Zhang} 

Assuming no fees, describing the new reserves in $t+1$ as the old reserves in $t$ plus/minus the traded amounts or $\Delta_A$ and $\Delta_B$ (trade asset B for asset A), it can be shown that (see also Appendix \ref{appendix:IntroAMM})
\begin{equation}
(R_A-\Delta_A )*(R_B+\Delta_B )=R_A*R_B\,. \label{eq:kfunc_long_main}
\end{equation}
Consequently, the amount of asset A received ($\Delta_A$) in exchange for a certain amount of asset B sent to the liquidity pool ($\Delta_B$) is defined by solving equation (\ref{eq:kfunc_long_main}) for $\Delta_A$ (see also Krishnamachari et al., 2021 for a general solution for more than two assets and unequal weights)\cite{krishnamachari}:
\begin{equation}
\Delta_A =\frac{R_A*\Delta_B}{R_B+\Delta_B}\,. 
\end{equation}
The resulting exchange rate offered by the liquidity pool for the demanded asset A ($\epsilon_{AMM}$) describes how much of asset A a trader will receive for a certain amount of B ($\Delta_A=\epsilon_{AMM}*\Delta_B$):
\begin{equation}
\epsilon_{AMM} = \frac{\Delta_A}{\Delta_B} =\frac{R_A}{R_B+\Delta_B}\,.
\end{equation}
This exchange rate of A ($\epsilon_A$) is a decreasing function of the amount of asset B ($\Delta_B$) to be exchanged. Consequently, trades with larger volumes will result in lower exchange rates; i.e., if traders or arbitrageurs want larger amounts of A, they have to send more units of B per unit of A to the liquidity pool. 

\begin{figure} [h]
\centering
\captionsetup{justification=centering}
\includegraphics[width=0.8\textwidth]{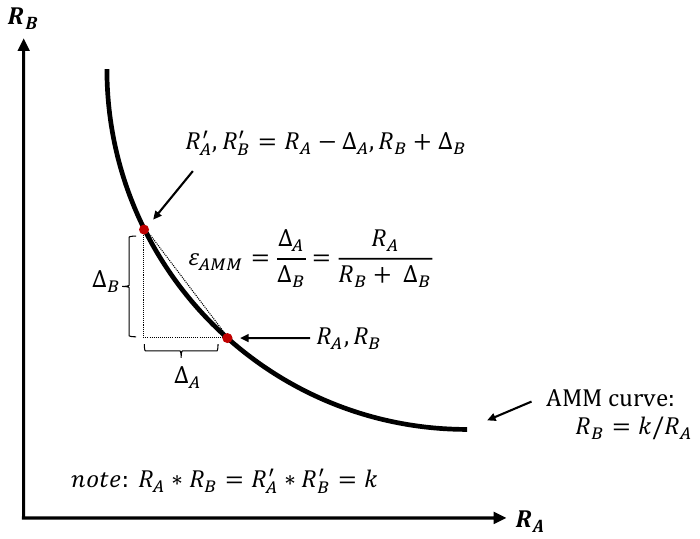}
 \caption{Illustration of the price function of a constant product function AMM.\cite{krishnamachari}}
 \label{graph5}
\end{figure}

Note that the reference price of asset A offered by the AMM ($p_{AMM}$) which describes how much of asset B a trader has to pay (send) to receive a certain amount of A ($\Delta_A*p_{AMM}=\Delta_B$) is the inverse of the exchange rate
\begin{equation}
p_{AMM}=\frac{1}{\epsilon_A} = \frac{R_B+\Delta_B}{R_A}\,. 
\end{equation}
Figure \ref{graph5} illustrates the exchange rates for a trade $\Delta_B$ $\rightarrow$ $\Delta_A$.

\vspace{12pt}

The curve ($R_B=k/R_A$) in Figure \ref{graph5} represents all combinations of reserves $R_A$ and $R_B$ in the liquidity pool for which $R_A*R_B=k$ holds. For every state of the reserves ${R_A,R_B}$ and any amount $\Delta_B$ to be exchanged, the curve describes how much the trader receives of the other asset $\Delta_A$.

Assume, for example, the liquidity pool consists of 50 units of A and 100 units of B. If a trader wants to receive 10 units of A in exchange for B, she has to send 25 units of B to the DEX. After this exchange, the liquidity pool consists of 125 units of B and 40 units of A, and the constant product function (\ref{eq:kfunc_long_main}) holds because $100*50=125*40$.
As every trade changes the reserves within the liquidity pool, the price changes with every trade. Since the price is defined as $\Delta_B/\Delta_A$, the spot price ($p_{AMM}$), which represents the price offered for an indefinite small amount, is defined by the slope of the constant product function\cite{krishnamachari}:
\begin{equation}
p_{AMM} = -\left( \frac{\partial R_B}{\partial R_A} \right) = \frac{k}{R_A^2}\,.
\end{equation}
Since $k=R_A*R_B$, the spot price is the ratio of the two reserves, i.e.,

\begin{equation}
p_{AMM}=R_B/R_A\,. 
\end{equation}
Assuming no fees, Angeris et al. (2021) show that the liquidity providers’ total relative gain (${\delta}^{LP}$) depends solely on the market price development\cite{Angeris} 
\begin{equation}
\delta^{LP}=\sqrt{p_m^T/p_m^1}\,. \label{eq:relgain}
\end{equation}
As a result, impermanent loss ($IL$) can formally be described as
\begin{equation}
IL=\frac{W_1*\delta^{LP}}{W_1*\delta^{Ref}}-1=\frac{2*\sqrt{p_m^T/p_m^1}}{1+p_m^T/p_m^1} -1 \label{eq:IL}
\end{equation}
with $\delta^{Ref}$ being the relative gain of the reference portfolio and $W_1$ the initial wealth of the liquidity provider.

According to (\ref{eq:IL}) it follows that without any fees a market price change will always result in an impermanent loss for the liquidity provider. Because liquidity providers earn fees for their services, their overall return has to take these fees into account. Since this overall return depends on complex interactions with arbitrageurs and traders, we develop an agent-based model to simulate these interactions.

\section{The Agent-Based Model} \label{sec:AMM_ABM}
We simulate the interactions between market prices, traders, arbitrageurs, liquidity providers, and the AMM of a liquidity pool for the WETH/USDC asset pair. This simulation is based on historical on-chain data collected from Uniswap V2 during the period from May 2021 to May 2022.\cite{Dune1}. Traders exchange WETH for USDC (or vice versa) with the liquidity pool. The AMM updates the exchange rate according to a constant product function as described in Section 2. Whenever the spot exchange rate determined by the AMM deviates from the market exchange rate, arbitrageurs will trade against the liquidity pool to earn risk-free profits. During the simulation, liquidity providers do not interact with the liquidity pool.

\subsection{General Procedure}
\begin{figure} [h]
\centering
\captionsetup{justification=centering}
\includegraphics[width=0.8\textwidth]{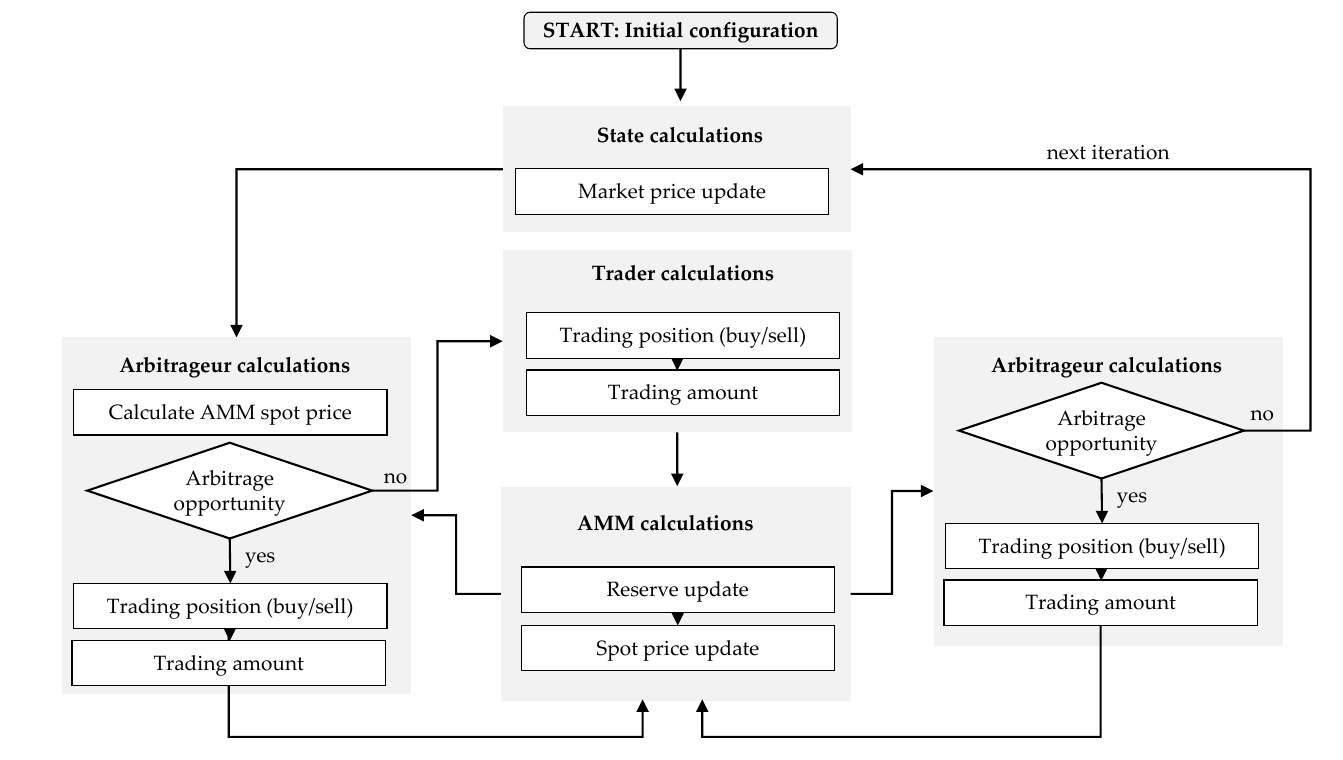}
 \caption{Process flowchart of the ABM.}
 \label{graph1}
\end{figure}

The simulation consists of the steps depicted in the flowchart of Figure \ref{graph1}. One loop represents one simulation period. This loop is repeated iteratively for a period of one year. Each loop consists of calculations which can be classified into four blocks: state calculations, arbitrageur calculations, trader calculations, and AMM calculations. Arbitrageurs will check whether there is an arbitrage opportunity after each state update and after each new trade. Therefore, individual calculation blocks can run multiple times within one iteration.

\vspace{12pt}

\subsection{Agent Decisions and other Calculations}

\textit{Initial configuration}—The simulation begins with an initial configuration that populates the “world” with representative traders and arbitrageurs as well as information on the asset market prices and the number of assets in the pool. We feed the initial configuration with effective values of Uniswap’s WETH/USDC pool (trades/fees/pool size) and Binance (market price) in 2021. This data comprises 1.3 million trades conducted over a period of one year (see Appendix \ref{appendix:Data} for more details). We have hereby excluded any MEV trades, since we account for arbitrage trades endogenously.\cite{Dune2,Etherscan}

\textit{State calculations}—After these initial configurations, the simulation process starts with the state calculations. First, market prices of the relevant assets are updated.

\textit{Arbitrageurs’ actions}—After this price update, arbitrageurs anticipate the AMM behavior and calculate Uniswap’s spot price (see also AMM calculations) and compare it with the market price. In the baseline simulation, we assume almost perfect market conditions for arbitrageurs; i.e., if spot prices deviate from market prices by more than the trading fee plus a small transaction cost, arbitrageurs will trade with the AMM. In particular, they will choose the amount that maximizes their profit and send it to the AMM. If, for example, the spot price for asset A is significantly lower than its market price, arbitrageurs will (i) buy asset B on a competing exchange, (ii) send it to the AMM, (iii) receive asset A, and (iv) sell it on a competing exchange. Thus, in the no-fee case, arbitrageurs’ profit function is
\begin{equation}
\pi=(p_A*\Delta_A-p_B* \Delta_B)\,.
\end{equation}
It can be shown that solving this maximization problem yields the optimal amount of $A$\cite{Angeris}
\begin{equation}
\Delta_{A}^{*}=(R_A-\sqrt{(R_A*R_B)/p_m})_+
\end{equation}
where $(x)_+ = max \{x,0\}$ for $x \in \mathbb{R}$. See also Appendix \ref{appendix:Maximization Problem} for the derivation and the solution including trading fees. Accordingly, arbitrageurs will trade in the opposite direction if the spot price of A at Uniswap is significantly above its (external) market price.

\textit{AMM calculations}—Whenever agents (arbitrageurs or traders) decide to trade, they send one of the two assets to the liquidity pool. These trades trigger AMM calculations according to the constant product function for two assets ($R_A*R_B=k$). Based on the asset type and amount provided by the traders or arbitrageurs, the AMM will calculate how much they will receive in return and how much will be captured as fees. In addition, due to the accumulated fees, the AMM will also update the constant $k$ for the next trade.

\textit{Traders’ actions}—In a next step, the trader summits an order to the AMM and exchanges an amount of A tokens ($-\Delta_{A}$) for B tokens ($\Delta_B$) or vice versa. This decision is based on the historical trading amount of Unsiwap's WETH/USDC pool. After traders’ actions and before the next iterations begins, a next round of arbitrageur calculations takes place.
\subsection{Experiments}
To analyze which factors impact the profits of liquidity providers, we conduct four experiments. First, we test if our model behaves as intended, by testing hypothesis regarding liquidity providers' profits in terms of trading activity and the size of the liquidity pool. Then, we test the impact of ETH price changes on liquidity providers’ profits by adjusting the ETH price growth. Finally, we test the impact of transaction costs and arbitrageur competition on their profits.

\textit{Trading activity}—An increase in trading volume increases accumulated trading fees over time, which should increase profits of liquidity providers. Thus, we adjust the trading volume by restricting the number of trades. More specifically, we create a subset that only consists of each \textit{n}th trade and then analyze the impact on profits.

\textit{Liquidity}—Liquidity providers’ income consists of fees from traders and arbitrageurs. Thus, the larger the liquidity pool, the lower the ratio between fees and liquidity, which reduces liquidity providers’ profits. Thus, we adjust the liquidity by increasing the initial pool size, i.e., increasing the initial WETH/USDC reserves.

\textit{Growth}—Literature predicts that without fees, a positive or negative trend in the market prices results in impermanent loss for the liquidity provider.\cite{AignerDhaliwal} Thus, we adjust the trend of the ETH price to analyze the impact on profit. More specifically, we simulate a price feed using geometric Brownian motion to match various expected growth rates $g$.\cite{Booker} In addition, we adjust traded quantities such that trading volumes remain unchanged:
\begin{equation}
p_t=p_{t-1}*e^{(g-0.5*\sigma^2 )*\Delta t+\sigma*\sqrt{\Delta t}*z_t}\text{.}
\end{equation}
\textit{Arbitrageur Transaction Costs}—High arbitrageur activity is generally believed to have a negative impact on liquidity providers’ profits. For example, Milionis et al. (2023) show that any arbitrage profits will result in losses for liquidity providers.\cite{Milionis} Arbitrageurs trade whenever spot prices deviate by more than the sum of fees and transaction costs. Thus, we adjust arbitrageurs’ transaction costs ($\tau$) --– which are a proxy for arbitrageur competition --– and analyze their impact on profits:
\begin{equation}
p_m>p_{AMM}*1/(1-fees-\tau)\,.
\end{equation}

\section{Results and Discussions} \label{sec:AMM_Results}
This section presents the results from the experiments in three steps. First, we replicate standard results regarding trading activity and pool size from previous studies. Second, we discuss the effect of different changes in market price. Finally, we analyze how competition among arbitrageurs affects liquidity providers’ profits. 

\subsection{Standard Results}
Based on the baseline scenario (see also Appendix \ref{appendix:Data}), we first test whether we can replicate standard results from previous literature, i.e., that an increase in trading activity or a decrease in the provided liquidity increases the profitability of liquidity providers. Thus, we run various simulations using adjusted trading volume and adjusted initial liquidity pool sizes. Unsurprisingly and in line with theory, the simulations show that higher trading activity results in more accumulated fees, which increases liquidity providers’ profits. In addition, an increase in liquidity decreases profitability because fees are distributed over a larger amount of invested liquidity. 
\subsection{Market Price Changes}
To understand the effect of the fee, we analyze the effect of market price changes on liquidity providers’ profits. We first analyze whether we can replicate results from literature that analyze impermanent loss under the no-fee assumption. We then conduct the same analysis but introduce protocol fees and compare results.

\begin{figure} 
\centering
\hspace{2cm}
\captionsetup{justification=centering}
\includegraphics[width=0.8\textwidth, trim=0cm 1cm 0cm 1cm, clip]{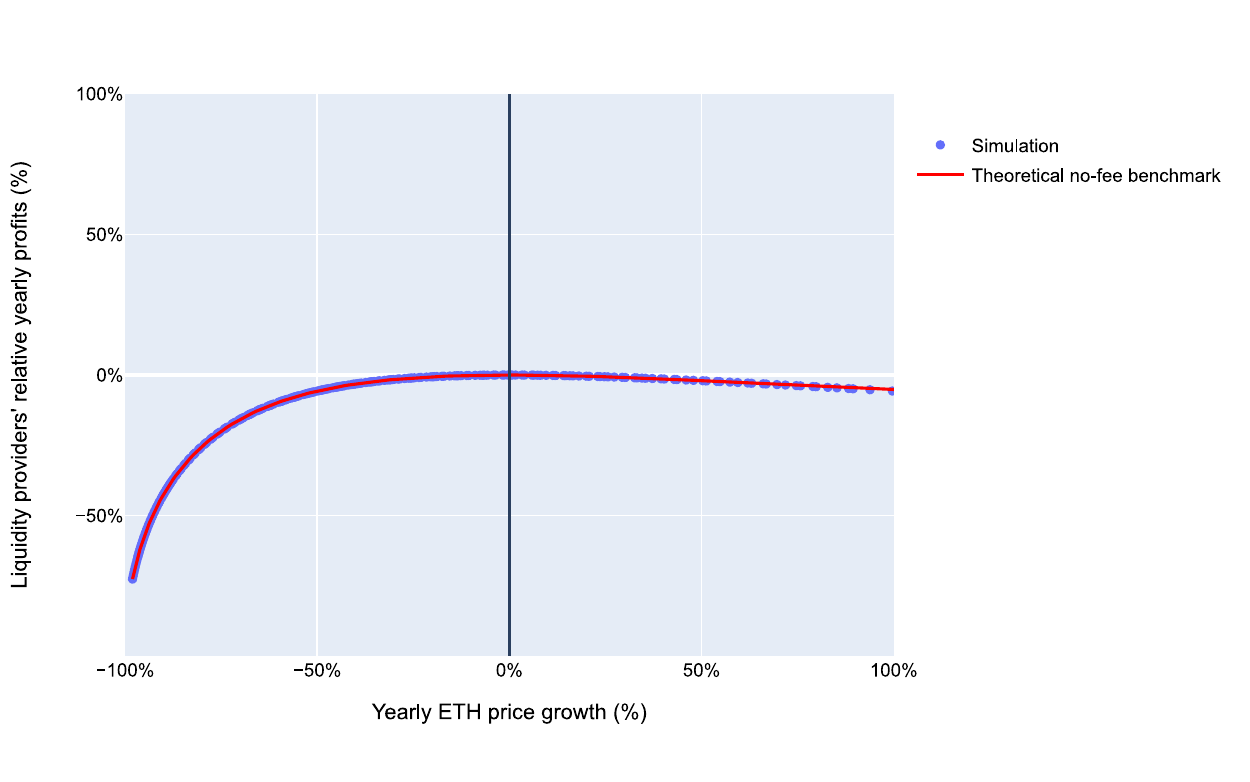} %see https://cloudconvert.com/ convert svg to png or pdf
 \caption{Simulation of liquidity providers' relative profits (without fees) assuming different market trends.}
 \label{no_fee_graph}
\end{figure}

\textit{No-fee experiment}—Based on the baseline scenario, we test whether we can replicate previous analysis that assumes fees to be zero.\cite{AignerDhaliwal} We can replicate this scenario by using our baseline scenario but assuming all fees and other transaction costs to be zero. Thus, we run various experiments assuming different market trends for the simulated ETH price path. More specifically, we vary the yearly price trend from -90\% (i.e., a price decrease from 2,765 USD to 277 USD) to +90\% (i.e., a price increase from 2,765 USD to 5,254 USD). For all the runs, we simulate liquidity providers’ profits. We find the impermanent loss figure described by several papers (see also Figure \ref{no_fee_graph}). Thus, the results are expected and in line with previous literature and theory.\cite{Angeris} However, as explained earlier, to get the full picture, it is essential to take into account fees as well

\textit{Fee experiment}—In this experiment, we analyze whether results change if we assume positive trading fees. Again, we run various experiments assuming different market trends for the simulated ETH price path as in the no-fee experiment. 

\vspace{12pt}

\begin{figure} [h]
\centering
\hspace{2cm}
\captionsetup{justification=centering}
\includegraphics[width=0.8\textwidth, trim=0cm 1cm 0cm 1cm, clip]{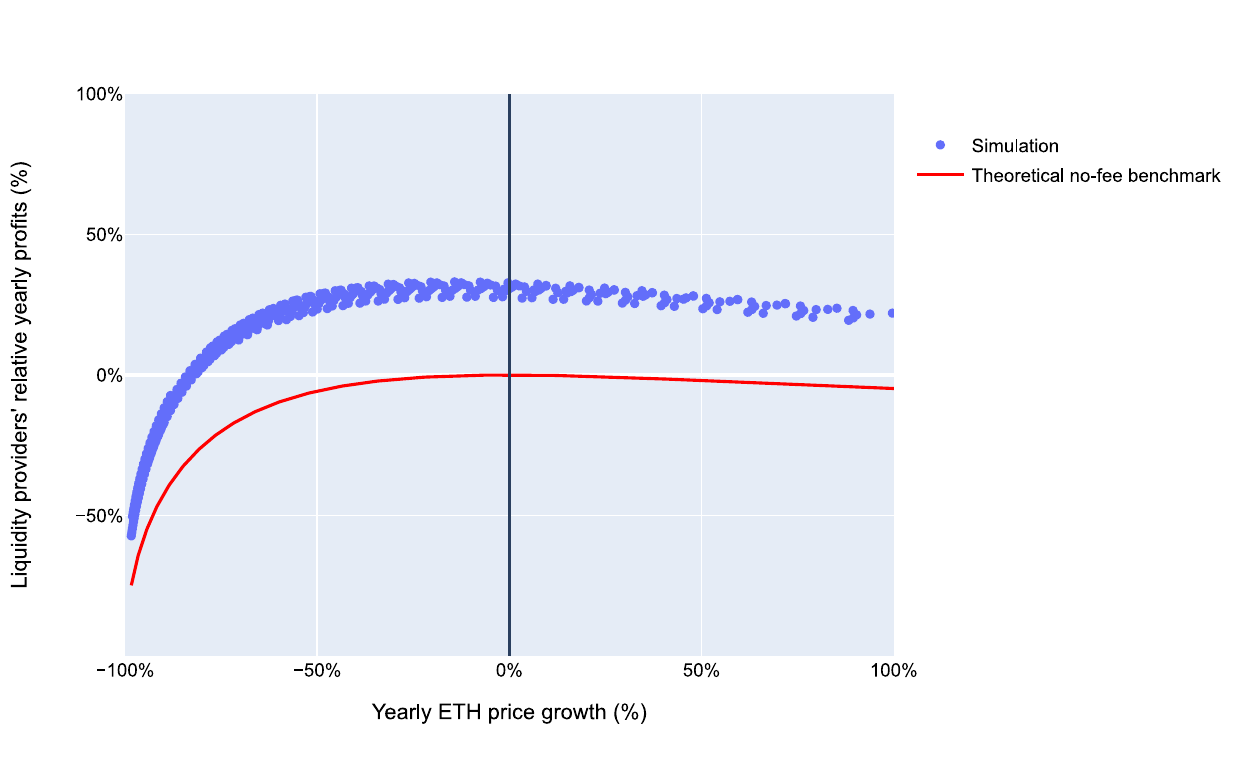} %see https://cloudconvert.com/ convert svg to png or pdf
 \caption{Simulation of liquidity providers' relative profits (with fees) assuming different market trends.}
 \label{graph2}
\end{figure}

Figure \ref{graph2} depicts the liquidity providers' returns (\textit{y}-axis) assuming different market price trends (\textit{x}-axis). According to the simulation, liquidity providers make a profit relative to a holding strategy as long as market prices do not drop by more than 75\% or increase by more than 300\% within a year. Thus, a decrease or an increase in the market price trend only has a significant effect on liquidity providers’ profits for exceptionally strong price movements.

The reason for this result is the impact of arbitrageurs and traders on the profits of liquidity providers: on one hand, arbitrageurs’ trades cause a change in the assets in the liquidity pool, leading to a drop in the profits of liquidity providers (the “rebalancing losses"). On the other hand, traders and arbitrageurs pay a fee for each trade, which increases liquidity providers’ profits (the “fee gains”). The overall result depends heavily on the size of these two effects: if the rebalancing losses are larger than the fee gains, then the net result will be negative (the liquidity providers earn less from fees than they lose from rebalancing). The fee effect itself depends strongly on the ratio between trading volume and liquidity (see also standard results, above). Assuming the ratio of our baseline model of around 2/100 (250 million USDC / 11.9 billion USDC) the rebalancing effect outweighs the fee effect only in case of exceptionally strong price movements.
\subsection{Arbitrageurs’ Transaction Cost Changes }
\begin{figure} [h]
\centering
\hspace{0.5cm}
\captionsetup{justification=centering}
\includegraphics[width=0.8\textwidth, trim=0cm 1cm 0cm 1cm, clip]{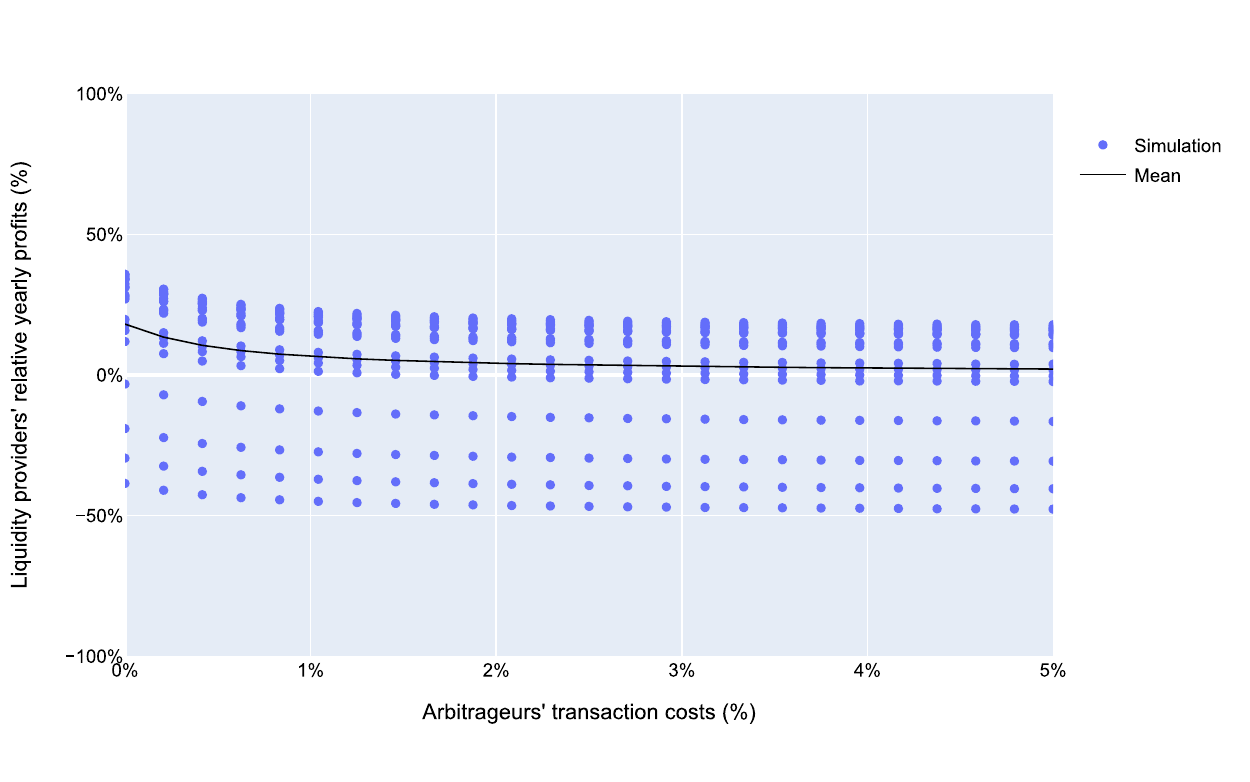}
 \caption{Simulation of liquidity providers' relative profits assuming different transaction costs for arbitrageurs.}
 \label{graph3}
\end{figure}

Finally, we analyze the impact of arbitrageurs' transaction costs on the returns of liquidity providers. It is commonly presumed that high arbitrageur activity adversely affects the profitability of liquidity providers. However, arbitrageurs also incur fees, causing a non-straightforward effect. We analyze the effect of arbitrageurs by varying their barriers to trade with the liquidity pool (i.e., their transactions costs). More specifically, we assume transaction costs from 0 to 5\% and simulate the profits of liquidity providers. The results are provided in Figure \ref{graph3}.

\vspace{12pt}

The findings are quite remarkable: our simulations reveal that as transaction costs for arbitrageurs rise, the profits of liquidity traders decrease. In other words, increased activity by arbitrageurs boosts the profits of liquidity providers, which contradicts the popular belief that arbitrage activity harms liquidity providers.
The dynamic nature of the setting accounts for this result. Although arbitrageurs reduce liquidity providers' profits through rebalancing losses, they also positively contribute by inducing fee gains. It can be shown that rebalancing losses are independent of the number of arbitrage trades and remain consistent over time as shown in equation (\ref{eq:relgain}). Conversely, gains from fees increase with arbitrage activity. Therefore, increasing arbitrageur activity increases liquidity providers' profits. This discovery differs from static analysis, since fee gains caused by arbitrageurs are often overlooked or modelled as a single trade. Our simulation shows that these fee by arbitrageurs are significant. Hence, it is vital to consider the fee gains from arbitrageurs in addition to rebalancing losses and trader fee gains to analyze liquidity providers’ profits.
\section{Conclusion} \label{sec:AMM_Conclusion}
In this paper, we investigate the impermanent loss problem in decentralized exchanges with automated market makers. We model the dynamic interactions between liquidity providers, traders, and arbitrageurs, and explicitly incorporate accumulated fees of traders and arbitrageurs into the profit function of liquidity providers. We use an agent-based model to simulate one of the most common AMM smart contract types (constant product AMM) and the behavior of their participants under various scenarios.

Our analysis demonstrates that impermanent losses are less severe than previously believed. While static analyses indicate that small price increases can result in losses for liquidity providers, our experiments, which incorporate dynamic effects from the accumulation of fees over time, suggest that price changes themselves do not necessarily lead to losses. Our analysis reveals that impermanent losses only occur when price increases are permanent, unexpected, and sufficiently large. A liquidity provider's return is positively influenced by trading fees and negatively impacted by (permanent) price changes and liquidity volume. In a competitive environment, liquidity provision is determined by expected price changes and trading volumes. Consequently, impermanent losses occur in the event of unexpected changes in these variables and if accumulated fees do not offset the losses. 

In addition, we show that the size of the impermanent loss decreases with the intensity of competition among arbitrageurs because fee gains from arbitrageurs are significant. Hence, it is vital to consider the fee gains from arbitrageurs in addition to rebalancing losses and trader fee gains. We conclude that developers of AMMs should prioritize fostering an arbitrage-friendly environment with low transaction fees, rather than attempting to prevent them. 

\newpage

%define the following sections to hide their Section Number (Notes Style)
\ledgernotes

%AUTHOR: comment out if using thebibliography
%\theendnotes

%AUTHOR: please read ledgerbib.bst usage notes by opening it in a text editor. We have modified it to include the use of the @misc item type for the proper formatting of online sources.

\bibliographystyle{ledgerbib}
\bibliography{output}

%AUTHOR: comment out, this is used to make sure the Creative Commons License
%image fits on page

\newpage 	
%define the following sections to have the Appendix Style

\appendix

\titleformat{\section}{\fontsize{13pt}{13pt}\normalfont\bfseries}{}{0em}{Appendix \Alph{section}:\quad}
\titlespacing*{\section}{0pt}{15pt}{7pt}

\section{Data}
\label{appendix:Data}

\vspace{12pt}

\begin{table}[h]
\centering
\small
\begin{tabular}{
  @{}
  l % First column as regular text column
  >{\centering\arraybackslash}p{42mm} % Center-aligned fixed width columns
  >{\centering\arraybackslash}p{42mm}
  @{}
}
\toprule
\textbf{Variable} & \textbf{Historical value} & \textbf{Initial value} \\
\midrule
Initial pool size & \O 240.6m USDC & 250m USDC \\
Trading volume per year (Q) & 11.9bn USDC & 11.9bn USDC (100\%) \\
Growth of the market price (g) & 0\% & 0\% \\
Market price return volatility ($\sigma$) & 1.2 & 1 \\
Number of trades (N) & 1.31m & 1.31m \\
Time period (T) & 1 year & 1 year \\
Protocol fee (fee) & 0.3\% & 0.3\% \\
\bottomrule
\end{tabular}
\caption{Historical data of our reference dataset and initial values.}
\label{AppendixTable}
\end{table}

\vspace{12pt}

\section{Introduction to Automated Market Makers}
\label{appendix:IntroAMM}

The generalized form of constant product AMM formulas is defined by 
\begin{equation}
\prod_{i=1}^{n} R_i^{w_i} = k\,,
\end{equation}
where $k$ is a constant value, $R_i$ are the reserves, i.e., the actual amount of asset $i$ in the liquidity pool, and $w_i$ is the weight that determines the ratio of the reserves in the pool.
For the purpose of simplicity, we focus on the simplest form with two assets and equal weights, which is also used by Uniswap V2. Note that our main results also hold for more complex functions. 
In line with previous research we describe the two token/asset constant product function by\cite{Zhang, Angeris}
\begin{equation}
R_A*R_B=k \label{eq:kfunc}\,.
\end{equation}
 In order that (\ref{eq:kfunc}) holds, each interaction of traders or arbitrageurs with the liquidity pool must, therefore, adjust the reserves $R_A$ and$R_B$ such that $k$ remains constant:
\begin{equation}
R_A^t*R_B^t=k\,,
\end{equation}
\begin{equation}
R_A^{t+1}*R_B^{t+1}=k\,. 
\end{equation}
Thus, a trade must satisfy
\begin{equation}
R_A^{t+1}*R_B^{t+1}=R_A^t*R_B^t\,. \label{eq:kfunctime}
\end{equation}
Assuming no fees, describing the new reserves in $t+1$ as the old reserves in $t$ plus or minus the traded amounts $\Delta_A$ and $\Delta_B$ (trade asset B for asset A), we can express (\ref{eq:kfunctime}) as
\begin{equation}
(R_A-\Delta_A )*(R_B+\Delta_B )=R_A*R_B\,.
\end{equation}

\vspace{12pt}

\section{Derivation of the Spot Market Price and Exchange Rate (Fees)}
\label{appendix:SpotWithFees}

Calculations including fees with $\gamma = 1-fee$:
\begin{equation}
(R_A-\Delta_A )*(R_B+\gamma*\Delta_B )=R_A*R_B \label{eg:kfunclongfee}
\end{equation}
Amount A
\begin{equation}
\Delta_A = \frac{R_A*\gamma*\Delta_B}{R_B+\gamma*\Delta_B}
\end{equation}
The resulting exchange rate offered by the liquidity pool for the demanded asset A ($\epsilon_{AMM}$), which describes how much of asset A a trader receives for a certain amount of B ($\Delta_A*\epsilon_{AMM}=\Delta_B$), is
\begin{equation}
\epsilon_{AMM} = \frac{\Delta_A}{\Delta_B} =\frac{\gamma*R_A}{R_B+\gamma*\Delta_B}\,.
\end{equation}
This exchange rate of A ($\epsilon_A$) is a decreasing function of the amount of asset type B ($\Delta_B$) to be exchanged. Consequently, trades with larger volumes result in a lower exchange rate; i.e., if traders or arbitrageurs want larger amounts of A, they have to send more units of B per unit of A to the liquidity pool.
Note that the reference price of asset A offered by the AMM ($p_{AMM}$), which describes how much of asset B a trader has to pay (send) to receive a certain amount of A ($\Delta_A*p_{AMM}= \Delta_B$), is the inverse of the exchange rate:
\begin{equation}
p_{AMM}=\frac{1}{\epsilon_A} = \frac{R_B+\gamma*\Delta_B}{\gamma*R_A}\,.
\end{equation}
For infinite small trades, this results in

Spot exchange rate:
\begin{equation}
\epsilon_{AMM}=\frac{\Delta_A}{\Delta_B} =\gamma*\frac{R_A}{R_B} 
\end{equation}

Spot price:
\begin{equation}
m_{AMM}=\frac{1}{\epsilon_A} = \frac{1}{\gamma} * \frac{R_B}{R_A} 
\end{equation}

\vspace{12pt}

\section{Derivation of the Arbitrageur's Maximization Problem}
\label{appendix:Maximization Problem}

If the market price for asset A ($p_A$) is significantly larger than its spot price $m_{AMM}$, arbitrageurs will (i) buy asset B ($\Delta_B$) on a competing exchange, (ii) send it to the AMM, (iii) receive asset A ($\Delta_A$), and (iv) sell it on the competing exchange. In this case, arbitrageurs profit function is

\begin{equation}
\pi=p_A*\Delta_A-p_B* \Delta_B\,. \label{eg:profitf}
\end{equation}

Denominated in the reference market prices, this results in the following maximization problem:

\begin{equation}
\begin{aligned}
\underset{\Delta_A}{\max} & \quad \pi = p_m * \Delta_A - \Delta_B \\
\text{subject to}& \quad (R_A - \Delta_A) * (R_B + \gamma * \Delta_B) = R_A * R_B, \\
& \quad \Delta_A, \Delta_B > 0\,. 
\end{aligned}
\end{equation}

In a \textbf{first step}, we put the constraint (AMM’s constant product function) into the maximization problem: to find the optimal $\Delta_A^*$ we solve AMM’s constant product formula (\ref{eg:kfunclongfee}) for $\Delta_B$ and put the solution (\ref{eg:profitf}) into the investors’ profit function (\ref{eq:DeltaB}). The resulting profit function depends solely on quantity $\Delta_A$ and not anymore on $\Delta_B$:

\begin{equation}
\Delta_B = \frac{1}{\gamma} * \left( \frac{R_A * R_B}{R_A - \Delta_A} - R_B \right)\,, \label{eq:DeltaB}
\end{equation}

\begin{equation}
\pi = p_m * \Delta_A - \frac{1}{\gamma} * \left( \frac{R_A * R_B}{R_A - \Delta_A} - R_B \right)\,.
\end{equation}
Note that in this format, the objective function resembles the problem of a profit-maximizing firm that operates in a competitive goods market selling good $\Delta_A$:
\begin{equation}
\pi(\Delta_A) = \text{Revenues}(\Delta_A) - \text{Costs}(\Delta_A)\
\end{equation}
with
\begin{equation}
Revenues = p_m*\Delta_A
\end{equation}
and
\begin{equation}
\text{Costs} = \frac{1}{\gamma} * \left( \frac{R_A * R_B}{R_A - \Delta_A} - R_B \right)\,.
\end{equation}
To optimize profits, investors seek the quantity $\Delta_A^*$ for which marginal revenues (MR) equal marginal costs (MC). 

Thus, in a \textbf{second step}, we calculate marginal costs and revenues
\begin{equation}
MR=MC\,,
\end{equation}

\begin{equation}
p_m = \frac{1}{\gamma} * \frac{R_A * R_B}{(R_A - \Delta_A)^2}\,. 
\end{equation}

In a \textbf{third step}, we solve for the optimal quantity $\Delta_A^*$:
\begin{equation}
\Delta_A = R_A \pm \sqrt{\frac{R_A * R_B}{\gamma * p_m}}\,.
\end{equation}
Solution for positive $\Delta_A$, $\Delta_B$, $R_A$, $R_B$, and $p_m$ 
\begin{equation}
\Delta_A^* = \left( R_A - \sqrt{\frac{R_A * R_B}{\gamma * p_m}} \right)
\end{equation}
and 
\begin{equation}
\Delta_B^{*} * \gamma = \sqrt{R_A * R_B * \gamma * p_m} - R_B\,.
\end{equation}
Furthermore, it can be shown that
\begin{equation}
\Delta_A^* = R_A * \left(1 - \sqrt{\frac{p_{AMM}}{\gamma * p_m}} \right)
\end{equation}
and
\begin{equation}
\Delta_B^* = \frac{1}{\gamma} * R_B * \left( \sqrt{\frac{\gamma * p_m}{p_{\text{AMM}}}} - 1 \right)\,.
\end{equation}
Note, if the AMM asks for a fee ($fee > 0$) arbitrageurs trade less and with smaller amounts. Including fees, arbitrageurs face larger marginal costs since they receive less for any amount sent to the protocol; see $1/\gamma$ in (\ref{eq:DeltaB}).

%\newpage
%here up^^

%\thispagestyle{pagelast}

%\theendnotes

\end{document}